\documentclass[%
frontmatterverbose, 
preprint,
showkeys,
 amsmath,amssymb,
 aps,
]{revtex4-1}

\usepackage[pdftex]{graphicx}
\usepackage{amssymb,amsfonts,amsmath}
\usepackage{color}
\usepackage[usenames,dvipsnames]{xcolor}
\usepackage{sidecap}

\definecolor{darkchocolate}{rgb}{0.55, 0.27, 0.07}

\begin{document}


\title{Excess vibrational modes of a crystal in an external non-affine field}

\author{Saswati Ganguly$^{\rm a}$ and Surajit Sengupta$^{\rm b*}$}
\affiliation{$^{\rm a}$Institut f\"ur Theoretische Physik II: Weiche Materie, Heinrich Heine-Universit\"at D\"usseldorf, Universit\"atsstra{\ss}e 1, 40225 D\"usseldorf, Germany.\\
$^{\rm b}$TIFR Centre for Interdisciplinary Sciences, 21 Brundavan Colony, Narsingi, Hyderabad 500075, India\\{\rm email:~surajit@tifrh.res.in}
}

\date{\today}


\begin{abstract}
Thermal displacement fluctuations in a crystal may  be classified as either ``affine" or ``non-affine". While the former couples to external stress with familiar consequences, the response of a crystal when {\em non-affine} displacements are enhanced using the thermodynamically conjugate field, is relatively less studied. We examine this using a simple model of a crystal in two dimensions for which analytical calculations are possible. Enhancing non-affine fluctuations destabilises the crystal. The population of small frequency phonon modes increases, with the phonon density of states shifting, as a whole, towards zero frequency. Even though the crystal is free of disorder, we observe growing length and time scales. Our results, which may have implications for the glass transition and structural phase transitions in solids, are compared to molecular dynamics simulations. Possibility of experimental verification of these results is also discussed.   
\end{abstract}

%
%
\maketitle

\section{Introduction}
While an ideal crystal~\cite{CL} is homogeneous at all length scales larger than the unit cell, a defining characteristic of glasses is the presence of structural and dynamic heterogeneities~\cite{hetero,glassbook,cavagna,biroli}. Such heterogeneities, cause ``soft spots"~\cite{manning}, participate in mechanical deformation~\cite{falk-review,falk} and contribute to localised transverse vibrational modes~\cite{shintani}. They have been argued to also give rise to phenomenon of ``boson peak"~\cite{laird,stz-boson1,stz-boson2, stz-boson3} viz. an enhancement in the weight of low frequency region in the phonon density of states. On the other hand, there are suggestions that the boson peak is unrelated to specific soft spots but simply represents an usual crystalline van Hove singularity broadened by disorder~\cite{boson,chuma-khao, eliott}. Indeed, explicit calculations for a model disordered solid shows~\cite{eliott} that hybridisation of crystalline vibrational states and a shifting of the lowest van Hove singularity to the low frequency region due to disorder contributes to the boson peak phenomenon. 

This paper is an attempt to reconcile these two, seemingly divergent viewpoints. Our argument proceeds as follows. Soft spots, detected in computer simulations~\cite{manning,falk} and in experiments~\cite{stz-boson2} on colloidal glasses undergo local structural rearrangements during shear deformation~\cite{schall}. These rearrangements are non-affine in the sense that they cannot be represented as an elastic strain on a reference volume~\cite{falk}. We have shown~\cite{sas1,sas2,sas3}, that thermal fluctuations can cause non-affine displacements even in a homogeneous crystal. In this case, such deformations are, however, transient and are predominantly defect (dislocation pair)~\cite{baluffi} precursors. We have also shown that it is possible to enhance (or suppress) these fluctuations in a crystal using an external field~\cite{sas2}. In this paper we ask what effect, if any, do these field enhanced non-affine deformations have on the phonon density of states of crystals? 

To answer this question we consider a very simple crystal consisting of particles connected by harmonic bonds and arranged in a two dimensional triangular lattice~\cite{harmdyn}. While we ignore anharmonic effects, we demonstrate below that the model is still capable of showing non-trivial behaviour. A great advantage of this model, on the other hand, is the fact that many results may be obtained analytically~\cite{sas1,sas2}, completely removing artefacts arising from systematic and random errors from computer simulations or experiments. 

Using the well defined projection formalism, which we have established in a previous publication~\cite{sas1}, we separate thermal particle displacements into affine and non-affine subspaces. An external field is introduced~\cite{sas2}, which couples selectively only to the non-affine component of the displacements. The phonon dispersion curve and the density of states is then obtained as a function of the field. As the field is increased, vibrational modes at the Brillouin zone boundary~\cite{ashcroft} becomes soft and the density of states gets enhanced in the low frequency region. The smallest van Hove singularity, which appears as a sharp peak also shifts to the low frequency region. As a corollary, we show that this shift is accompanied by a many-fold increase in the correlation length and relaxation time of non-affine displacements. 

Our results demonstrate, for what we believe to be the first time, that non-affine fluctuations are the primary cause for the boson peak phenomenon. These fluctuations are predominant in a disordered solid but can also be artificially tuned in an ordered crystal which, nevertheless remains perfectly homogeneous over large length and time scales. Disorder can broaden this response, which, again, we show explicitly by introducing random interactions in our model. 

The rest of the paper is organised as follows. In the next section we explain the projection formalism for obtaining non-affine displacements and define our model system and simulation procedure to test our results. This is followed by a description of our main (analytic) results which are compared to simulations. We end the paper with a discussion on the role of disorder. 

%
%

\section{Non-affine fluctuations} 
An ideal crystal at a temperature $T \neq 0$ is a rather well understood object - its physics being completely specified by lattice vibrations~\cite{ashcroft}.  The magnitude of lattice vibrations is often used as a measure to determine the stability of a crystal and its melting properties~\cite{charu}. 

 For the special case of a {\it harmonic} crystal, defined, for example by the Hamiltonian, 
\begin{equation} 
{\cal H}_0 = \sum^N_{i=1} \frac{{\bf p}_i^2}{2 m} + 
\frac{K}{2} \sum_{i=1}^N\sum_{j\in\Omega,i<j} 
(|{\bf r}_j - {\bf r}_i| - |{\bf R}_j - {\bf R}_i|)^2 
\label{hamil0}
\end{equation}
many essential thermodynamic quantities may be obtained exactly. Here ${\bf p}_i$ the momentum, $m$ the mass, ${\bf r}_i$ the instantaneous
position, and ${\bf R}_i$ the reference position of particle $i$. The length scale is set by the lattice parameter $l$, the energy scale by $K l^2$, and the time scale by $\sqrt{m/K}$. We have chosen $l = m = K = 1$. A dimensionless inverse temperature is given by $\beta = K l^2/k_B T$, with $k_B$ the Boltzmann constant. The interactions have a range equal to the size of $\Omega$ - a typical coordination volume. 

In Refs.~\cite{sas1} and ~\cite{sas2} we have described in detail how particle displacements in such a system may be projected onto affine strains and non-affine displacement components. While ~\cite{sas1} gave the general theory and specific examples of the one dimensional chain and the two dimensional triangular lattice, ~\cite{sas3} extended this formalism to the two dimensional honey-comb structure. Below, we give a brief summary of this procedure for completeness. While our procedure is perfectly general and can be carried through for any lattice in any dimension, we present results only for the two dimensional ($d=2$) triangular lattice with $\Omega$ restricted to the first neighbour shell consisting of a central particle and its six neighbours. 

Within $\Omega$, around particle $i$, define relative atomic displacements, ${\bf \Delta}_{j} = {\bf u}_j-{\bf u}_i$ with $j\neq i \in\Omega$. To obtain the best fit local affine deformation ${\mathsf D}$, one minimises  $\sum_j [{\bf \Delta}_{j} - {\mathsf D}({\bf R}_{j} - {\bf R}_{i})]^2$ with the non-affinity parameter $\chi({\bf R}_i) > 0$ given by the minimum value of this quantity~\cite{falk}. In Ref.~\cite{sas1} we showed that this minimisation procedure is equivalent to a 
projection of ${\bf \Delta}_i$ onto a non-affine subspace defined by the projection operators ${\mathsf P}$, where $\chi({\bf
R}_i)= {\bf \Delta}^{\rm T}{\mathsf P}{\bf \Delta}$ and $\Delta$ is a column vector constructed out of the  ${\bf \Delta}_i$. The projection operator ${\mathsf P}^2 = {\mathsf P} = {\mathsf I}-{\mathsf R}({\mathsf R}^{\rm T}{\mathsf R})^{-1}{\mathsf R}^{\rm T}$ and ${\mathsf R}_{j\alpha,\gamma\gamma^{\prime}} = \delta_{\alpha\gamma}R_{j\gamma^{\prime}}$, with the central particle $i$ taken to be at the origin.

In~\cite{sas1}, we showed how the statistics of $\chi$ may be obtained for (\ref{hamil0}). For instance, the thermal average $\langle \chi \rangle$ is given by the sum of the eigenvalues, $\sum_\mu \sigma_\mu$, of ${\mathsf P}{\mathsf C}{\mathsf P}$ with $C_{i\alpha,j\gamma} = \langle \Delta_{i\alpha}\Delta_{j\gamma}\rangle$, the covariance matrix of the {\em coarse-grained} displacements. Using rather standard statistical mechanical
techniques, using appropriate generating functions, one can obtain~\cite{sas1,sas2} explicitly both the single point probability, $P(\chi)$, and two point, spatio-temporal correlation function $$C_{\chi}({\bf R},t) = \frac{\langle\chi(0,0)\chi({\bf R},t)\rangle-\langle\chi\rangle^2}{\langle\chi^2\rangle-\langle\chi\rangle^2}.$$
These results are accurate for any solid at low temperatures and high densities where the harmonic approximation, used in deriving them, is valid.

The thermal average $\langle \chi \rangle$ for a crystalline solid is non-zero at any finite temperature, it is also, on the average, exponentially correlated in space and time with the spatial and temporal scales being set by the lattice parameter and the microscopic relaxation time. However, as we show below
 it is possible to enhance the expectation value and correlation of $\chi$. We use a device introduced in~\cite{sas2} for this purpose.  
 
Consider then the  extended Hamiltonian, 
 \begin{equation} 
{\cal H} = {\cal H}_0 - h_X \sum_i^N\sum_{jk\in \Omega} ({\bf u}_j-{\bf u}_i)^{\rm T}{\bf P}_{j-i,k-i}({\bf u}_k-{\bf u}_i).
\label{hamil1}
\end{equation}
The term multiplying $h_X$ involves the elements of the projection operator ${\bf P}_{ij}$, a function only of the reference configuration, that acts
on particle displacements. Note that the extra term in (\ref{hamil1}) can also be represented as $h_X X$ where $X=\sum^N_i \chi({\bf R}_i)$, the spatial average of the non-affine parameter $\chi$. A positive value of the non-affine field $h_X$ enhances non-affine distortions of the lattice~\cite{sas1,sas2}, namely, fluctuations in particle positions projected onto a subspace spanned by the eigenvectors associated with the non-affine eigenvalues $\sigma_\mu$. 

\begin{figure}[h!]
\begin{center}
\includegraphics[width=0.5\textwidth]{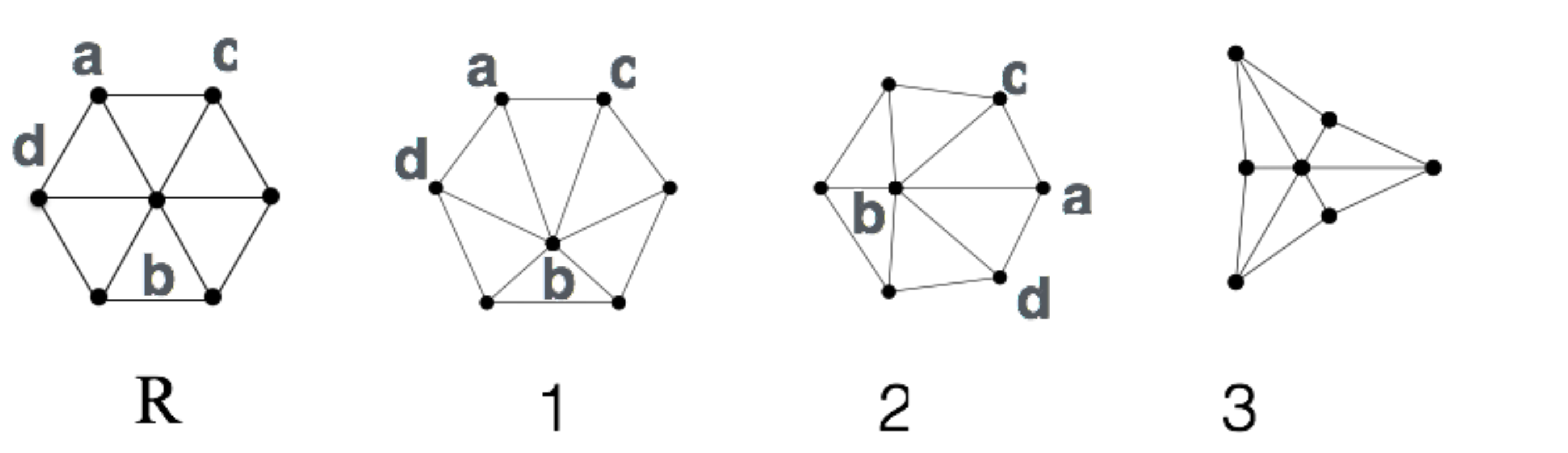}
\caption{Non-affine distortions of the reference configuration ${\bf R}$. The
configurations $1$ and $2$ represent (degenerate) eigenvectors of
${\mathsf P}{\mathsf C}{\mathsf P}^{\rm T}$ corresponding to the
two largest eigenvalues $\sigma_1$ and $\sigma_2$. The next largest
eigenvalue belongs to the eigenvector $\mu = 3$. Note that $\mu = 1$
and $2$ represent distortions where nearest neighbours $a, b$ move away
from each other while next nearest neighbours $c,d$ come closer tending
to nucleate a dislocation dipole~\cite{sas2}. The third eigenvector
represents coming together of $4$ particles to form a triangular lattice
of tetramers with twice the lattice parameter.}
\label{naf-modes}
\end{center}
\end{figure}

Taking $\Omega$ as the nearest neighbour shell consisting of particle
$i$ and its $6$ neighbours, we obtain $12$ possible lattice modes. There
are $d \times d = 4$ affine distortions in two dimensions, viz.~local
volume, uniaxial and shear distortion of $\Omega$ together with local
rotations. The remaining $8$ modes are non-affine and appear as non-trivial
eigenvectors of  ${\mathsf P}{\mathsf C}{\mathsf P}$. The
distribution of the eigenvalues is not uniform with the lowest eigenvalues
being separated from the rest by a large gap~\cite{sas2,sas3}. In
Fig.~\ref{naf-modes}, we have plotted three of the softest non-affine modes
which have the highest eigenvalues and therefore make the largest contribution to $\chi$. Note that the reciprocal of the eigenvalue is proportional
to the curvature of the Hessian projected in the direction of the
eigenvector~\cite{sas2}. The two largest eigen-distortions, ($\mu = 1,2$)
are degenerate and have been shown~\cite{sas2} to represent fluctuations
which tend to nucleate a dislocation dipole. The third eigenvector
($\mu = 3$), with an eigenvalue about five times smaller, represents a
fluctuation where the original triangular lattice transforms to another
similar triangular structure consisting of tetramers with double the
lattice parameter -- similar to a period doubling transition commonly observed in
one dimension~\cite{CL}. The remaining higher energy modes represent
more complicated distortions of $\Omega$~\cite{sas1,sas2}.  

Analytic calculations of correlation functions for the model solid
presented in the next section have been carried out using procedures
explained in detail in Refs.~\cite{sas1} and \cite{sas2} and will not
be repeated here. 

Simulation results report molecular dynamics (MD) calculations~\cite{frenkel}. MD simulations in the canonical ensemble, i.e.~at constant particle number $N$, volume $V$ and temperature $T$, were done using a leapfrog algorithm, coupling the system to a Brown and Clarke thermostat~\cite{frenkel}. The size of the systems ranges from from $N=100$ to $N=40000$ particles. Typically, unless otherwise stated, we used an MD time step of $\delta t = 0.002$ in reduced time units and reduced inverse temperature $\beta=100$.

\section{Results}
Increasing $h_X$ increases $\langle \chi \rangle$ locally. Most of this contribution comes from the softest non-affine modes, corresponding to $\mu = 1,2$ (see Fig.~\ref{naf-modes}). Note that any linear combinations of these modes are also degenerate and so the direction of the modulation wavenumber, with
magnitude $\chi^{-1/2}$, varies in space pointing randomly in all directions consistent with crystal symmetry. This modulation is incompatible with global, crystalline order and therefore remains a localised distortion of the lattice. We now study the consequences of enhancing the probability of {\em all} non-affine modes using an external field. 

An advantage of studying the properties of a solid using a harmonic approximation is the possibility of obtaining exact results {\em even
in the presence of the non-affine field} $h_X$. Indeed, the complete statistical mechanics of this system can be obtained analytically (apart from numerical integrals and Fourier transforms)~\cite{sas1,sas2,sas3} as long as the periodic crystalline phase is stable. In order to understand how crystal properties are affected by the non-affine field, $h_X$, we calculate the Hessian ${\cal D}({\bf R},{\bf R}') = \partial^2 {\cal H}/\partial {\bf u}({\bf R})\partial {\bf u}({\bf
R}')$. A Fourier transform then yields the dynamical matrix, whose eigenvalues are the phonon dispersion curves. 

For  convenience of representation we write the dynamical matrix as a sum of two terms ${\mathcal D}({\bf q})={\mathcal D}_{\rm harm}+{\mathcal D}_{X}$, where the first term ${\mathcal D}_{\rm harm}$ corresponds to the purely harmonic part of (\ref{hamil1}) and the second term, ${\mathcal D}_{X}$ corresponds to the interaction arising due to the nonaffine field ($= -h_{X}NX$). The harmonic part of the dynamical matrix~\cite{harmdyn} is given by,
%
\begin{eqnarray}
{\mathcal D}_{0} & = & k
\left( \begin{array}{cc}
3-2\,\cos(q_{x})-\cos(\frac{1}{2}q_{x})\cos(\frac{\surd3}{2}q_{y}) & 
\sqrt{3}\sin(\frac{1}{2}q_{x})\sin(\frac{\surd3}{2}q_{y})\\
\sqrt{3}\sin(\frac{1}{2}q_{x})\sin(\frac{\surd3}{2}q_{y}) & 
3[1-\cos(\frac{1}{2}q_{x})\cos(\frac{\surd3}{2}q_{y})]\\
\end{array}
\right) \nonumber \\
& = &k\left( \begin{array}{cc}
3-2\mathcal{A}_{1}-\mathcal{A}_{2}&\sqrt{3}\mathcal{A}_{3} \\
\sqrt{3}\mathcal{A}_{3} & 3-3\mathcal{A}_{2}\end{array}\right), 
\end{eqnarray}
%
where $k$ and $l$ are the spring constant and the lattice parameter
of the reference harmonic lattice with $q_{x}$ and $q_{y}$ being the
orthogonal components of the wave vector ${\bf q}$. The second term of
the dynamical matrix, $\mathcal{D}_{X}$
can be derived from the global non-affinity parameter $X$. The individual
components of $\mathcal{D}_{X}$ may be represented as
\begin{align*} 
\mathcal{D}_{X}^{\nu\mu}=-h_XN\sum_{j}
\frac{\partial^{2}X}{\partial {u_{\mu j}}\partial {u_{\nu i}}}e^{-iq.({\bf R}_{j}-{\bf R}_{i})}.
\end{align*} 
Here ${\bf R}_{j}$ represents the lattice positions of all the
neighbouring particles lying within the interaction volume of the
particle with reference ${\bf R}_{i}$ and ${u_{\mu i}}$ with $\mu=x,y$
represents the $x$ and $y$ components of the displacement vectors of the
particles. Taking double derivatives of $X$ and performing the sum over
$j$ leads to the final expression for ${\mathcal D}_{X}$,
\[ \mathcal{D}_{X}=-2\,h_X\,{\mathcal A}_{X}\left( \begin{array}{cc}
1& 0\\
0& 1\end{array}\right)\] 
with  
\begin{eqnarray}
\mathcal{A}_{X}&=&\frac{1}{6}\Bigg\{ 
\left[60-28\cos(q_{x}l)-56\cos\Big(\frac{q_{x}l}{2}\Big)
\cos\Big(\frac{q_{y}\sqrt{3}l}{2}\Big) \right] \nonumber\\
&& + \left[4\cos(2q_{x}l)+8\cos(q_{x}l)\cos(q_{y}\sqrt{3}l)\right] 
\nonumber \\
&& + \left[8\cos\Big(\frac{q_{x}3l}{2}\Big)
\cos\Big(\frac{q_{y}\sqrt{3}l}{2}\Big)+4\cos(q_{y}\sqrt{3}l)\right]
\Bigg\}. \nonumber 
\end{eqnarray}
Note that the leading order term in $\mathcal{A}_{X}(q)$ is of order
${\mathcal O}(q^4)$, so that $h_X$ does not contribute to the speed of
sound or to elastic constants.

Assembling all terms, the full dynamical matrix in the presence of the
nonaffine field is therefore given by,
\begin{equation*} 
\mathcal{D}= k \left( \begin{array}{cc}
3 -2 \mathcal{A}_{1}- \mathcal{A}_{2} - 
\frac{2 h_X}{k}\mathcal{A}_{X}&\sqrt{3} \mathcal{A}_{3} \\
\sqrt{3} \mathcal{A}_{3} & 3 -3 \mathcal{A}_{2} - 
\frac{2 h_X}{k}\mathcal{A}_{X}\end{array}\right)
\end{equation*}

\begin{figure}[ht]
\begin{center}
\includegraphics[width=0.7\textwidth]{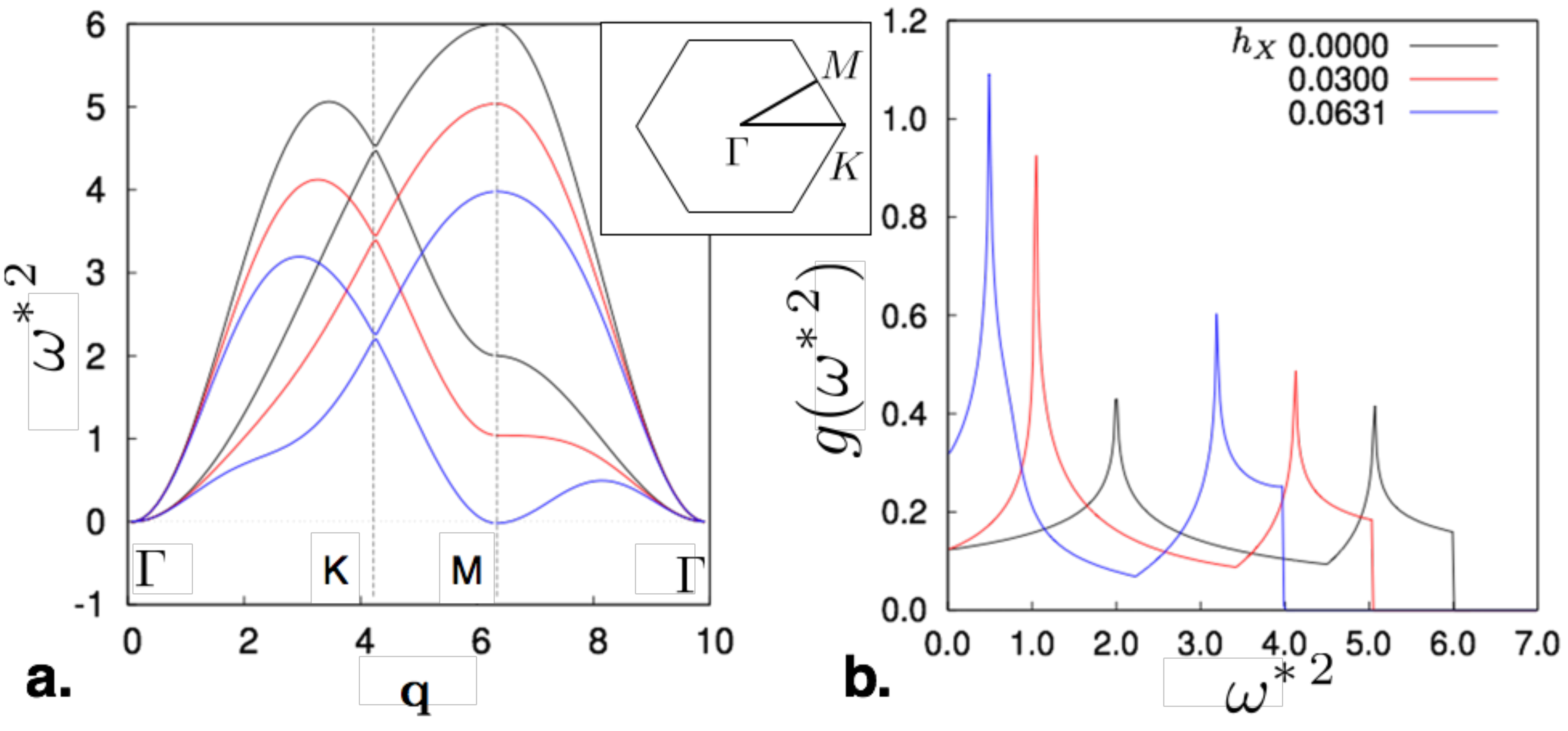}
\caption{{\bf a.} The phonon dispersion curves showing the square of the reduced frequency $\omega^{*2} = m\omega^2/K$ as a function of the wave number ${\bf q}$ and {\bf b.} Scaled DOS
calculated for three different values of $h_X$ indexed
by the same colours. Note the softening of some modes
in the dispersion curves which is reflected in the increase of the weight
in the DOS for $\omega \to 0$ as $h_X$ increases.   }
\label{dis_dos}    
\end{center}
\end{figure}

The phonon dispersion curves shown in Fig~\ref{dis_dos} are obtained by
diagonalising $\mathcal{D}$. The eigenvalues, which correspond to the
phonon frequencies $\omega(q)$, are given by,
\begin{eqnarray}
\omega^2(q)& = &3k-k\mathcal{A}_{1}(q)-
2k\mathcal{A}_{2}(q)-2h_X\mathcal{A}_{X}(q) \nonumber \\
& & \pm k\left[(\mathcal{A}_{1}(q)-\mathcal{A}_{2}(q))^{2}+
3\mathcal{A}_{3}^{2}(q)\right]^{\frac{1}{2}}.\nonumber
\end{eqnarray}
The phonon density of states is then evaluated numerically in the usual
manner~\cite{ashcroft} by counting the number of $q$ values within the
Brillouin zone which correspond to the same value of $\omega$.
The analytically obtained phonon dispersion curves $\omega_s({\bf q})$
are shown in Fig.~\ref{dis_dos}{\bf a}. The limiting behaviour
of $\omega_s({\bf q})$ as $q \to 0$ {\em is independent of $h_X$}. This
implies that, to {\color{black}leading} order, $h_X$ does not change any of the mechanical
properties of the crystalline solid, nor does it introduce any stresses
while enhancing non-affinity.

Increasing $h_{X}$ leads to softening of certain phonon modes. At
small ${\bf q}$, the transverse phonon modes are softened showing
that the solid becomes nearly unstable to large wavelength shear modes. At
large ${\bf q}$ near the zone boundary at the high symmetry $M$ point,
the phonon mode becomes soft and $\omega({\bf q}) \to 0$ as $h_X$ is
increased. Inspection of the eigenvectors associated with this mode
reveals that these are identical to the non-affine mode $\mu = 3$
(see Fig.~\ref{naf-modes}). The phonon density
of states (DOS) $g(\omega) = \sum_s\int_{BZ} d{\bf
q}\,\omega_s({\bf q})\,\delta[\omega - \omega_s({\bf q})]$ is shown in
Fig.~\ref{dis_dos}{\bf b} for the same values of $h_X$.  The DOS shows a
prominent shift towards $\omega = 0$ as $h_{X}$ increases. 

A similar shift of the DOS to small $\omega$ is a well-known consequence of introducing disorder in crystals~\cite{boson}. Note that in our calculations, crystal periodicity and homogeneity is imposed. Therefore it is surprising that a similar effect is seen even in the absence of disorder. This supports our contention that this shift is due to the enhancement of non-affine fluctuations. Disorder only makes these fluctuations more predominant. 

\begin{figure}[h!]
\begin{center}
\includegraphics[width=0.7\textwidth]{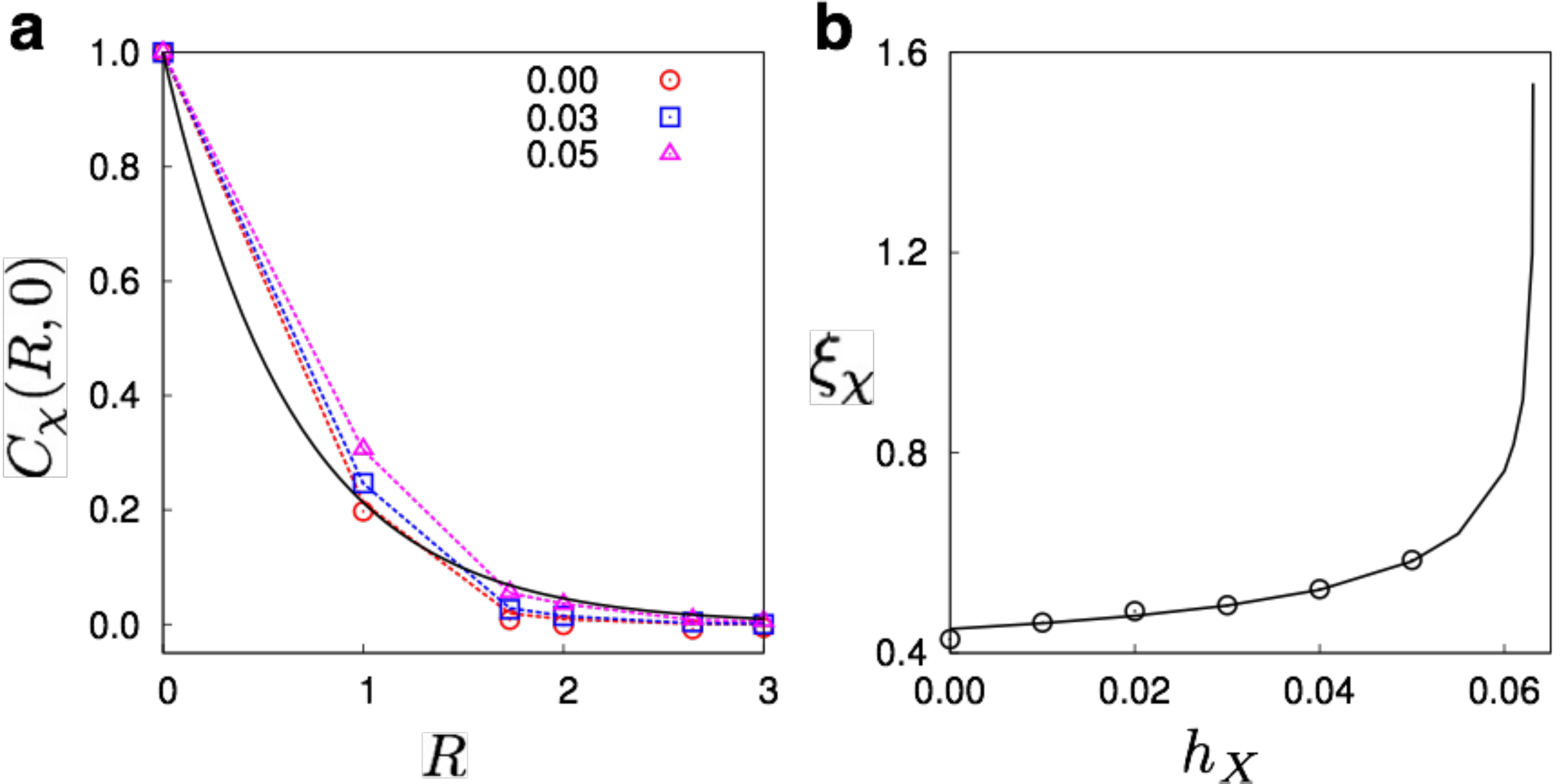}
\caption{\label{spatial} 
{\bf a.} Shows the
radially averaged, equal time, spatial correlations, $C_\chi(R,0)$.  The points are simulation results for $h_X = 0., 0.03$ and $0.05$ and the 
lines are from analytic calculations. {\bf b.} Correlation lengths $\xi_\chi$ are obtained for both cases by fitting an exponential
decay to the initial part of the curves (black curve, shown for $h_X = 0$). {\bf b.} Shows the result for $\xi_\chi$ as a function of $h_X$: points simulation results and curve from analytic theory. Note the divergence as $h_X$ increases.
}
\end{center}
\end{figure}

\begin{figure}[h!]
\begin{center}
\includegraphics[width=0.7\textwidth]{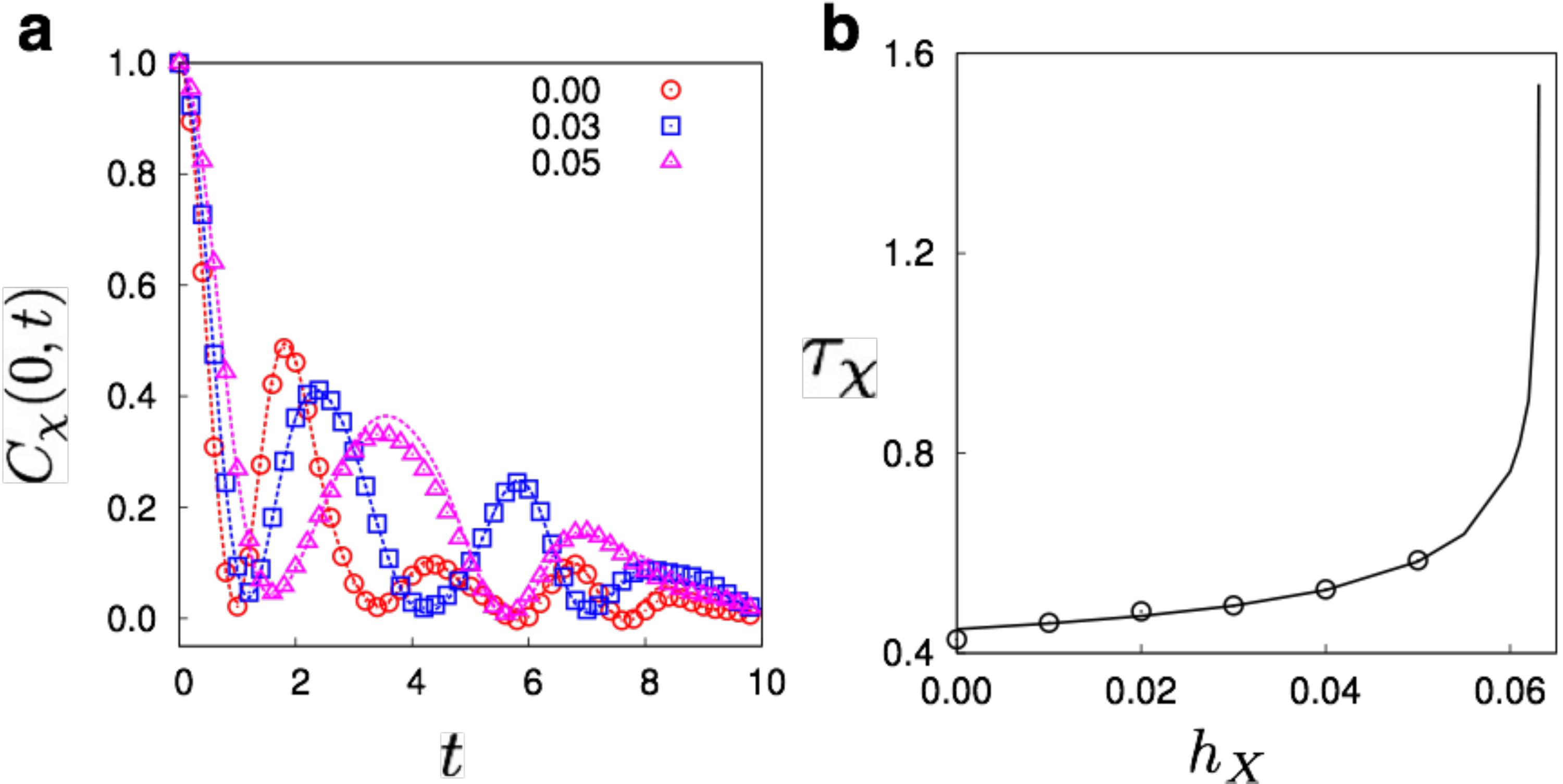}
\caption{\label{time} 
{\bf a.} Temporal correlation $C_\chi(0,t)$ for the same values of $h_X$. In this case the correlation time $\tau_\chi$ is the time at which $C_\chi(0,t)$ decreases to half its initial value. {\bf b.} The divergence of $\tau_\chi$: lines and points have same meaning as in Fig.~\ref{spatial}.
}
\end{center}
\end{figure}

In Figs.~\ref{spatial} and \ref{time} we compare analytic results for the spatial non-affine correlation function with those obtained from MD calculations. The spatio-temporal correlation functions were obtained analytically as in~\cite{sas2}. For each of these correlation functions we have obtained correlation lengths $\xi_\chi$ and times $\tau_\chi$, which are also shown. For the values of $h_X$ where MD results corresponding to the homogeneous crystal are available, our calculations show excellent agreement with simulations. At larger values of $h_X$, both $\xi_\chi$ and $\tau_\chi$ shows a divergence. Indeed all $\sigma_\mu^{-1}$ also vanish as $\sqrt{h_X - h_\mu^0}$ pointing to an underlying saddle-node bifurcation point beyond which a crystalline solid cannot exist. Simulations show that somewhat before this point is reached, however, the solid becomes heterogeneous~\cite{pleat}. 


\section{Discussion and conclusion}
In this paper we show that enhancing non-affine fluctuations alone using an external field is sufficient to produce an excess of low frequency vibrational modes in a crystal {\em without} disorder. This occurs due to a shift of the lowest van-Hove singularity of the crystal to low frequency. These results are quite similar to the findings of Ref.~\cite{eliott} where a similar shift was observed as an effect of disorder. In our case, the solid, on the other hand, remains ordered, homogeneity and periodicity being imposed for all our analytic calculations. We thus conclude that disorder is {\em a sufficient but not necessary condition} for the appearance of boson peak phenomena. Non-affine displacements, which cause a shift of the van-Hove singularity, can be enhanced either by making the crystal disordered or independently using a specific external field.   

What happens when we, additionally, introduce disorder? We present results for our model solid using a simple disorder model introduced in Ref.~\cite{dilu} which is also similar in spirit with that of Ref.~\cite{eliott}. In this model, the reference configuration is maintained with the disorder appearing only in the spring constants of the bonds such that $K = 1 + \delta K$. Here, $\delta K$ is taken as a random variable distributed uniformly in the range $(-K_1,K_1)$. We find that even this seemingly innocent looking disorder, when sufficiently large, is enough to make drastic changes in the structure of the non-affine modes. 
\begin{figure}[t]
\begin{center}
\includegraphics[width=0.5\textwidth]{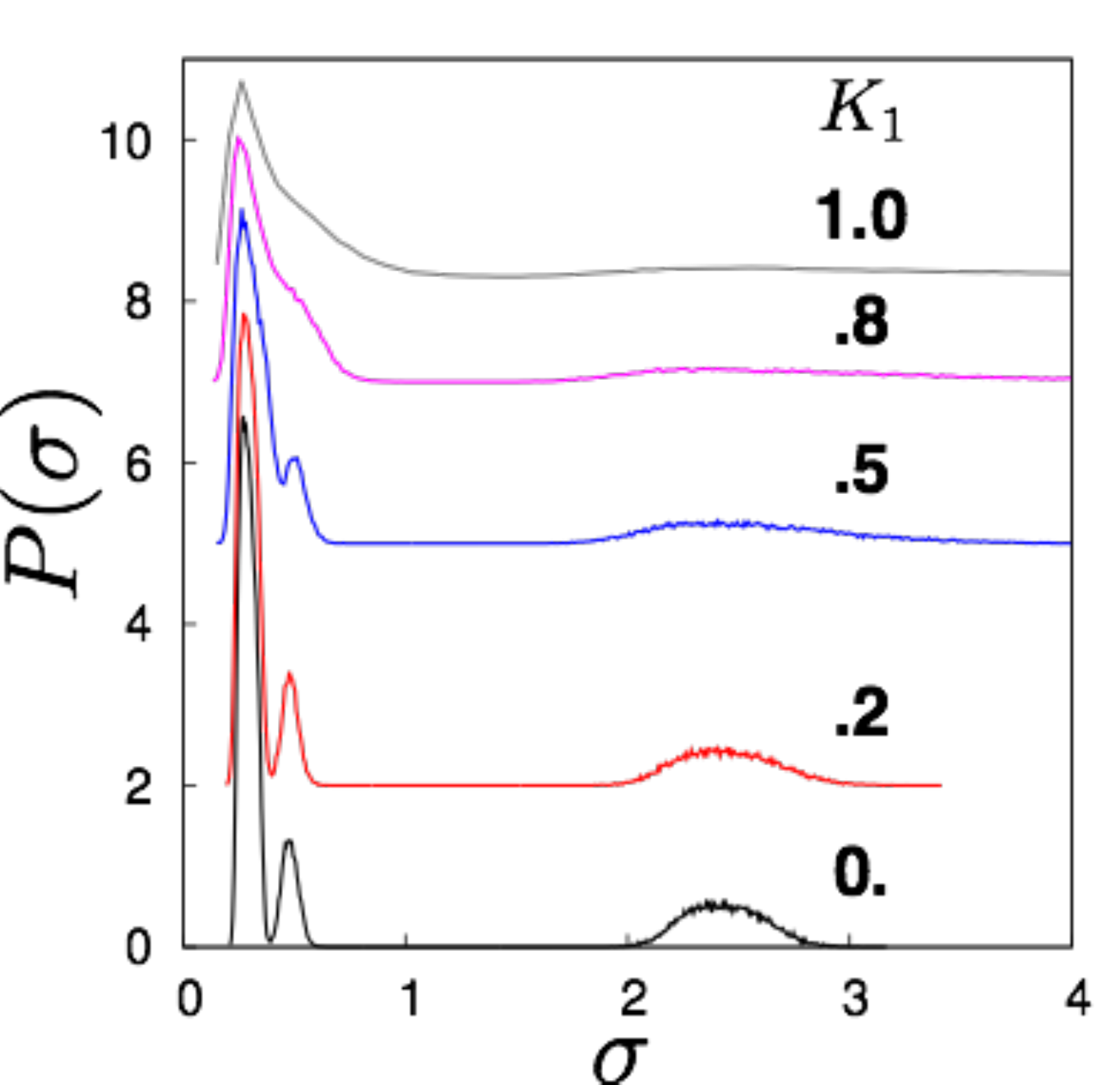}
\caption{\label{disorder} 
Plot of the probability distribution $P(\sigma)$ of the eigenvalues of ${\mathsf P}{\mathsf C}{\mathsf P}$ at $h_X = 0$ and for various values of disorder $K_1$ as defined in Ref.\cite{dilu}. For the pure solid ($K_1 = 0$) the distribution has peaks centred at each $\sigma = \sigma_\mu$ with only thermal broadening. As disorder is increased, the peaks broaden and the identity of each separate non-affine mode is lost.  
}
\end{center}
\end{figure}

This is illustrated in Fig.~\ref{disorder} where we have plotted the probability for obtaining an eigenvalue $\sigma$ for a local ${\mathsf P}{\mathsf C}{\mathsf P}$ matrix computed using the particle coordinates obtained from our MD simulations. For zero disorder this shows a series of well defined, but thermally broadened peaks, centred around $\sigma = \sigma_\mu$ - the eigenvalues obtained analytically for the homogeneous and periodic lattice. The peak for the largest $\sigma_\mu$ corresponds to dislocation precursors~\cite{sas2} and is separated from the rest of the structure by a large gap where there are no eigenvalues. As disorder increases, non-affine excitations loose their distinction and the mode structure becomes more and more diffuse. Each local region now contributes a different set of vibrational modes thereby broadening the density of states as seen in ~\cite{chuma-khao}.  

The external field  $h_X$, nevertheless remains, at present, a mathematical device realisable only in computer simulations. However, we have mentioned in Ref.~\cite{sas2} how one may be able to generate such a field experimentally in a colloidal solid. Briefly, one can use video microscopy to obtain information about the current position of atoms and selectively apply appropriate forces using dynamic laser traps. In case such a feedback system is actually set up in the future, we believe that the results shown here and earlier~\cite{sas2} would be valuable as a benchmark. In the future we would like to extend this work to higher dimensions and towards more realistic materials.

\begin{acknowledgments}
One of the authors (SS) remembers and acknowledges many interesting discussions with Charusita Chakravarty during the preliminary stages of this work. The promise of a more extensive engagement and possible collaboration with her on this subject was tragically cut short by her ill-health and passing away. The authors also thank J. Horbach, P. Sollich and S. Karmakar for discussions. Travel support from the FP7-PEOPLE-2013-IRSES grant no: 612707, DIONICOS is gratefully acknowledged. SG acknowledges support from  CSIR Senior Research Fellowship.
\end{acknowledgments}
%
%


\end{document}